\documentclass[%
aip-cp,
 jmp,%
 amsmath,amssymb,
preprint,%
]{revtex4-1}

\usepackage{graphicx}

\usepackage{dcolumn}%
\usepackage{bm}%
\usepackage{neuralnetwork}
\usepackage{tikz}

\def\ltsima{$\; \buildrel < \over \sim \;$}
\def\gtsima{$\; \buildrel > \over \sim \;$}
\def\simlt{\lower.5ex\hbox{\ltsima}}
\def\simgt{\lower.5ex\hbox{\gtsima}}

\usepackage{color}

\begin{document}
\preprint{ }

\title{Solving stiff ordinary differential equations using physics informed neural networks (PINNs): simple recipes to improve training of vanilla-PINNs}%

 
\author{Hubert Baty$ $  \\ {Observatoire Astronomique,
 Université de Strasbourg, 67000 Strasbourg, France \\ 
 hubert.baty@unistra.fr
  \quad \\ }}

\date{\today}%

\begin{abstract}
Physics informed neural networks (PINNs) are nowadays used as efficient machine learning methods for solving differential equations. However,
vanilla-PINNs fail to learn complex problems as ones involving stiff ordinary differential equations (ODEs). This is the case of
some initial value problems (IVPs) when the amount of training data is too small and/or the integration interval (for the variable like the time) is too large. 
We propose very simple recipes to improve the training process in cases where only prior knowledge at initial time of training data
is known for IVPs. For example, more physics can be easily embedded in the loss function in problems for which the total energy is conserved.
A better definition of the training data loss taking into account all the initial conditions can be done.
In a progressive learning approach, it is also possible to use a growing time interval with a moving grid (of collocation points) where the differential equation residual is minimized.
These improvements are also shown to be efficient in PINNs modelling for solving boundary value problems (BVPs) as
for the high Reynolds steady-state solution of advection-diffusion equation.
\end{abstract}


\maketitle

\clearpage

\section{Introduction}

The use of neural networks (NNs) to solve differential equations is revisited in a tutorial paper (see Baty \& Baty 2023 and references therein). Basically, a
classical NN can give a non linear approximation of the whole desired solution by using a dataset of known particular values. This is a supervised
learning method which consists in finding a mapping function between given inputs values (for example the time variable) and their output values (for example
the solution). This dataset is used to parameterize the NN such that it minimizes the error between solution predicted by the NN and true known solution from the dataset
during a training procedure. The convergence is achieved by minimizing a loss function which expression is based on error estimate, using for example the mean squared error.
Finding ``good'' parameters is achieved by solving an optimization problem using a gradient descent algorithm that relies on automatic differentiation to back-propagate
 gradients through the network (Baydin et al. 2018).

However, the amount of available known solution data (so called training data) is in general very small. These are typically the initial
or boundary values in problems involving ordinary differential equations (ODEs). In these cases, NN cannot be used as it is a bad extrapolation tool.
An approach called physics-informed neural networks (PINNs), has been thus proposed in order to tackle the limitations of classical NN (Raissi et al. 2017, 2019). The basic idea
is to provide additional information corresponding to the physics. The method consists in evaluating the solution at some other set of data points
(called collocation points) at which the equation residual is minimized.  A second loss function corresponding to the physics is thus defined and added
to the previous one in the learning process. The training is penalized by this additional constraint and the space of available solutions is thus restricted, being partly
driven by the original data and also partly driven by the physics. 

The use of PINNs to solve ODEs has been clearly illustrated in Paper 1 (Baty \& Baty 2023). In particular, it has been shown that data knowledge
representing only the initial conditions can be sufficient when the equations are weakly non linear. However, this not the case for strongly
non linear stiff problems (as for Van Der Pol oscillator for example). As a consequence, the use of PINNs in the latter cases
can be successful under the condition that a larger set of training data is used. The training procedure can be improved by adding another physical
information like the energy conservation (i.e. an additional constraint) in problems for which it is effectively conserved, but such limitations called failure modes in the literature remain
(Krishnapriyan et al. 2021). Note that we focus on ODEs in this study for the sake of simplification, but the problems discussed in this paper also concern the
integration of partial differential equations (PDE's). These difficulties are even worse when one to integrate over a rather large time interval.
Many improvements of these so-called vanilla-PINNs have been proposed in the literature, that are often based on self adaptive procedures
(of the collocation points, activation function, etc.). However, none of these methods appears to be effective on all the equations of concern (Karniadakis et al. 2021,
Xiang et al. 2022).
In this work, we propose very simple recipes that can be easily implemented with the vanilla-PINNs in order to ameliorate the training procedure.
The improved results obtained in this study are illustrated by directly comparing to some benchmark results shown in Paper 1.

The paper is organized as follows. In Section 2, we give a short summary of the PINNs technique. 
Two recipes based on modifying the loss function in two ways are investigated in Section 3.
In Section 4, we investigate a third recipe where the grid of collocation points is modified along with the progression
of the training process. Finally, conclusions are drawn in Section 5.

\section{Physics-Informed Neural Networks}

\subsection{The basics of neural networks for ODEs}

We first consider the desired solution $u(t)$ of an ODE (see below) with ${u_\theta} \simeq u$ being the approximated solution
at different $t$ values, where $\theta$ is a set of model parameters. For IVPs, the variable $t$ would generally be a time parameter and should be replaced
by a spatial coordinate parameter for boundary value problems (BVPs). Using a classical neural network approximating the desired solution, we can write,
\begin{equation}
{u_\theta} ( t) =  ( \mathcal{N}^L \circ \mathcal{N}^{L-1} ...\  \mathcal{N}^0) ( t) ,
\end{equation}
where the operator $\circ$ denotes the composition and $\theta =  \lbrace \boldsymbol{W}^l,  \boldsymbol{b}^l  \rbrace_{l=1,L}$ represents the trainable parameters
(with weight matrices and bias vectors) of the network (see Paper 1 for more details on the $ \mathcal{N}^l$ functions).
The network architecture schematized in Figure 1, is organized in $L+1$ layers with neurons connected in adjacent layers. 
A single input layer containing the input variables $t$ is connected to $L - 1$ hidden layers (two layers with four neurons in the schematic example of Figure 1), and finally to an output layer for the solution
${u_\theta}$. The goal is to calibrate its parameters $\theta$ such that ${u_\theta} $ approximates the target solution $u(t)$.
An activation function is also necessary in order to to introduce non-linearity into the output of each neuron.
In this work, the most commonly used hyperbolic tangent $tanh$ function is chosen.

The optimization problem is based on the minimization of a loss fonction that can be expressed as,
\begin{equation}
   L_{data} (\theta) = \frac  {1} {N_{data} } \sum_{i=1}^{N_{data} } \left| {u_\theta} (t_i ) - u_{data}^i  \right|^2 ,
\end{equation}
where a set of $N_{data}$ data is assumed to be available for the known solution at different times $t_i$ that are called the training data ($ i = 1, N_{data}$), which includes
the initial and/or boundary conditions.

  \begin{figure}[!t]
\centering
 \includegraphics[scale=0.3]{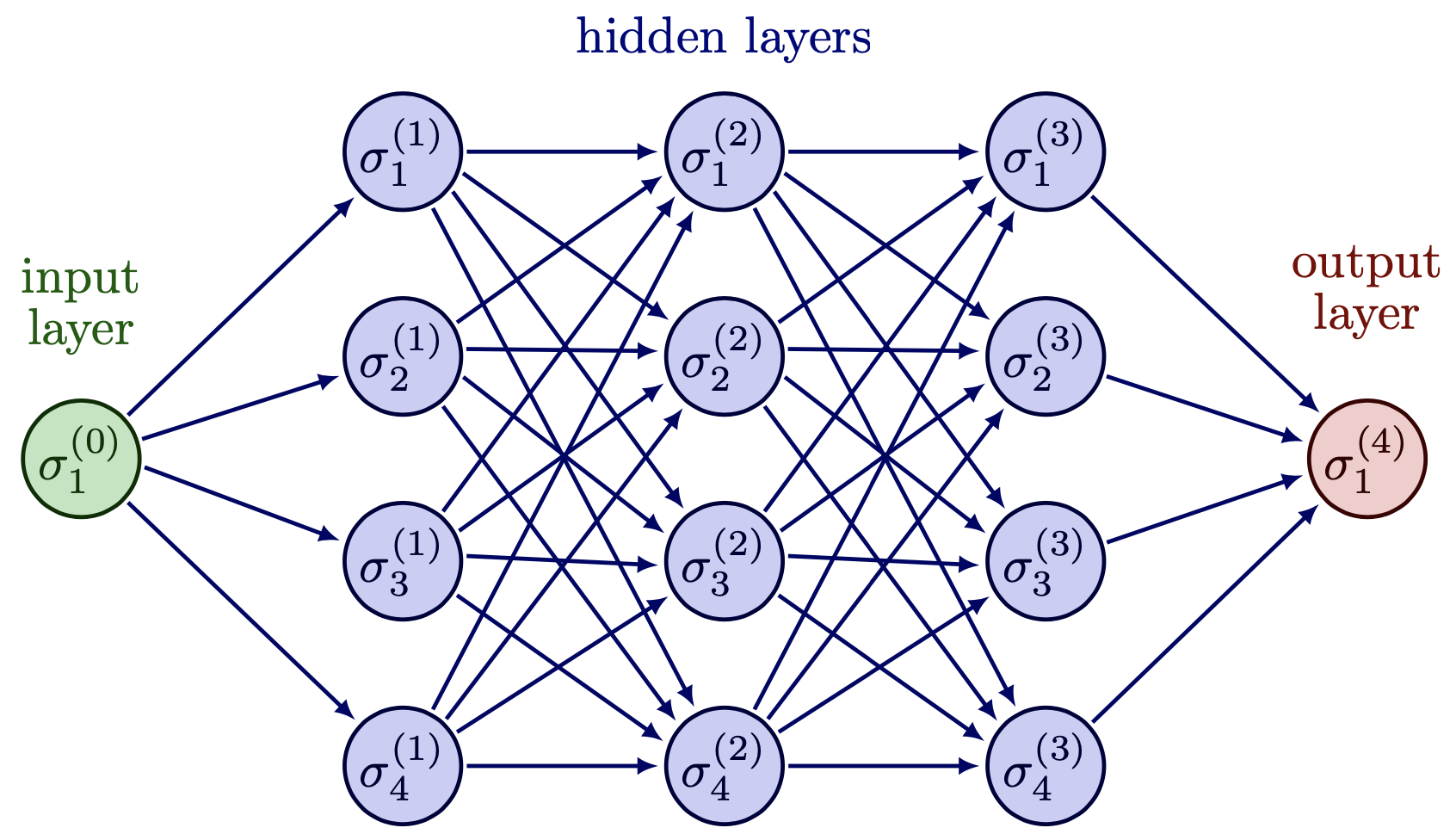}
  \caption{Schematic representation of a structure example for a standard NN. The input layer has one input variable (i.e. one neuron noted $\sigma_1^0$) representing
  for example a time coordinate. Three hidden layers with four neurons per layer are connected with the input and the output layer
  (one neuron noted $\sigma_1^4$), where the latter has a single variable (one neuron) representing the desired solution
  ${u_\theta}$.
      }
\label{fig1}
\end{figure}

\subsection{The basics of vanilla-PINNs for ODEs}

 \begin{figure}[!t]
\centering
 \includegraphics[scale=0.3]{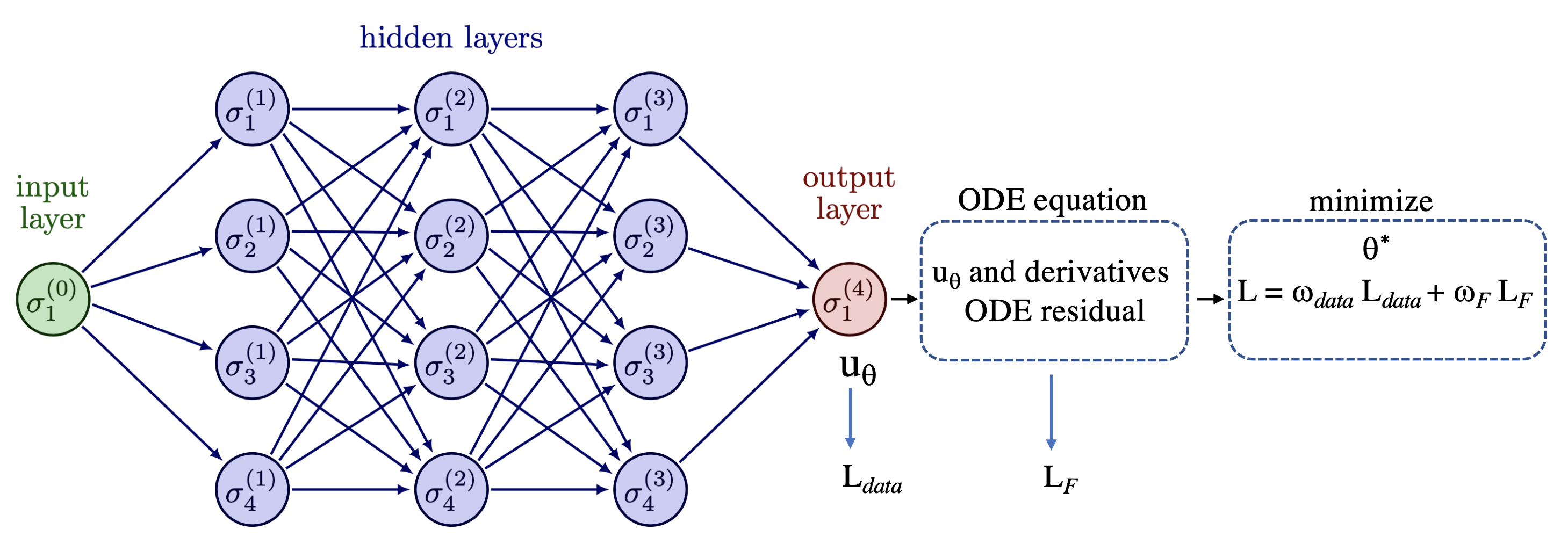}
  \caption{Schematic representation of the structure for a vanilla-PINN modelling an ODE solution.
  The previous NN architecture (see previous figure) is used to evaluate
  the residual of the ODE equation (via ${u_\theta}$ and corresponding derivatives). The two partial loss functions ($L_{data}$ and $L_{ \mathcal{F}}$)
 are used to form a total loss function with associated weights (see text) that is finally minimized.
       }
\label{fig2}
\end{figure}

Let us now introduce an ODE in the following residual form
\begin{equation}
   \mathcal{F} \left( t, u,   \frac  {du} {dt}, \frac  {d^2u} {dt^2},... \right) = 0,   \ \ \ \ \ \     t \in  \left[ t_0,T \right] ,
\end{equation}
with imposed initial conditions and/or boundary conditions depending on the problem considered. The exact number of initial/boundary conditions necessary to solve
the equation obviously depends on the order of the equation. Note that in case of an ODE with an order $n$ higher or equal to two, an equivalent system of $n$
equations can be also used (see below). In the original form, the  basics of vanilla-PINNs is based on the use of a second loss function defined as
\begin{equation}
   L_{ \mathcal{F}} (\theta) = \frac  {1} {N_c} \sum_{i=1}^{N_c}  \left| \mathcal{F} [ {u_\theta} (t_i )  ] \right|^2 ,
\end{equation}
that must be evaluated on a set of $N_c$ points generally called collocation points that are not necessarily coinciding with the training data points.
Note that one can evaluate exactly the differential operators at the collocation points in  $L_{ \mathcal{F}}$
and $\mathcal{F}$ by using automatic differentiation. This same technique of automatic differentiation is also used to compute derivatives with respect to the network weights (i.e. $\theta$),
that is necessary to implement the optimization procedure (see below). Note that contrary to the use of
standard numerical schemes, the derivatives can be obtained at machine precision. In this work, we use Pytorch Python open source software libraries
facilitating thus the latter operations.

A composite total loss function can be consequently formed as
\begin{equation}
              L  (\theta)    =   \omega_{data} L_{data} (\theta)  + \omega_{\mathcal{F}}L_{ \mathcal{F}} (\theta),
\end{equation}
where an optimal choice of values for hyper-parameters $(\omega_{data},  \omega_{\mathcal{F}})$ allow to ameliorate the eventual unbalance between
the partial losses during the training process. These weights can be user-specified or automatically tuned. In the present work, for simplicity we fix the $\omega_{data} $
value to be constant and equal to unity, and the other weight parameters including $ \omega_{\mathcal{F}}$ are determined with values varying from case to case.
A gradient descent algorithm is used until convergence towards the minimum is obtained for a predefined accuracy (or a given maximum iteration number) as
\begin{equation}
             \theta^{k+1} =  \theta^{k} - \eta  \nabla_{\theta}     L  (\theta^k) ,
\end{equation}
for the $k$-th iteration also called epoch in the literature,
leading to $  \theta^{*}  = \operatorname*{argmin}_\theta  L  (\theta)$, where $\eta$ is known as the learning rate parameter.
In this work, we choose the well known $Adam$ optimizer. A standard automatic differentiation technique is necessary to compute derivatives (i.e. $\nabla_{\theta}$) with respect to
the NN parameters (e.g. weights and biases) of the model (Raissi et al. 2019). A schematic representation of the vanilla-PINNs is shown in Figure 2. Note that
in this schematic figure, a single input neuron representing time or space coordinate for ODEs must be replaced by two neurons for a partial differential equation
having spatio-temporal ($x, t$) dependences.
Moreover, in cases where a set of $n$ differential equations is considered, the output neuron must be replaced by $n$ neurons associated with the
$n$ solution variables that need to be learned.

\section{Modifying the loss fonction}
\label{setup}

\subsection{Adding a constraint and associated partial loss function based on energy conservation}
In Paper 1, different benchmark tests are investigated mainly based on second order differential equations like the harmonic oscillator, non linear pendulum,
and anharmonic oscillators. In these cases, the total energy $E$ is conserved and fully determined by the initial conditions. It is thus possible
to add a corresponding additional constraint, with an associated third loss function called $L_E$. This constraint can be written in a residual form,
$E - E_0 = 0$, with $E_0$ being the total energy at $t = 0$. The latter residual form and associated loss function $L_E$ are thus evaluated
at the collocation points in a similar way as for the residual equation and loss function $L_{ \mathcal{F}}$. Consequently, one gets
\begin{equation}
   L_E (\theta) = \frac  {1} {N_c} \sum_{i=1}^{N_c}  \left| E (t_i ) - E_0  \right|^2 ,
\end{equation}
The total loss function is consequently modified by
incorporating a new term weighted  by a new hyper-parameter $\omega_E$,
\begin{equation}
 L  (\theta)    =   \omega_{data} L_{data} (\theta)  + \omega_{\mathcal{F}}L_{ \mathcal{F}} (\theta) + \omega_{E} L_{E} (\theta) .
 \end{equation}
It has been shown that the results are considerably ameliorated when compared
to a case without this additional constraint. For example, this is illustrated in Figure 8 of Paper 1 for the harmonic oscillator problem.

\subsection{Using an hybrid loss function for data }

\begin{figure}[!t]
\centering
 \includegraphics[scale=0.4]{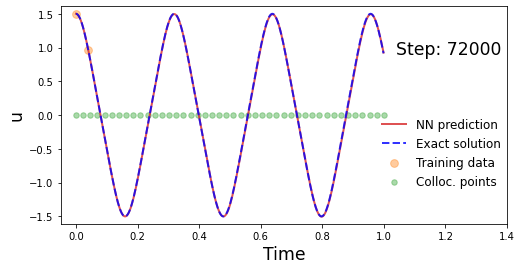}
 \includegraphics[scale=0.4]{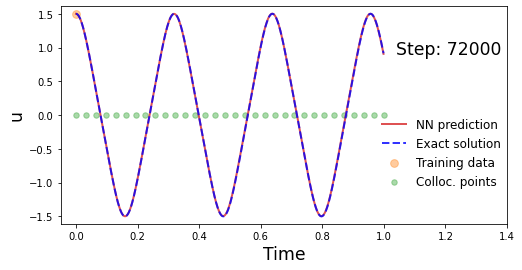}
  \caption{PINN solutions for the anharmonic oscillator (see text). Left and right panels show cases using, two training data without the hybrid data loss
  function and one training data with the hybrid data loss, respectively. The exact solution is obtained by a classical Runge-Kutta integration (using a method
  of order $4$ with $1000$ uniform time-steps.
       }
\label{fig3}
\end{figure}   

\begin{figure}[!t]
\centering
 \includegraphics[scale=0.5]{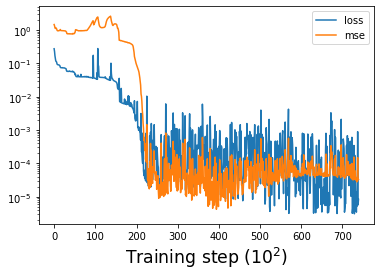}
 \includegraphics[scale=0.5]{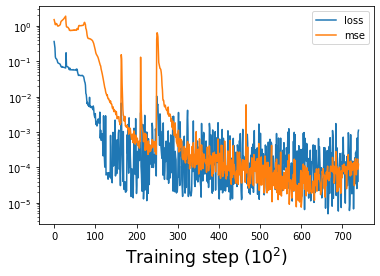}
  \caption{Histories of the total loss function $L$ and $MSE$ corresponding to the two cases of the previous figure respectively. The $MSE$ is evaluated using the standard
  expression, $MSE =   \frac  {1} {N_{eval} } \sum_{i=1}^{N_{eval} } \left| {u_\theta} (t_i) - u_{eval}^i  \right|^2 $, where the evaluation ${u_\theta} (t_i)$ is done
  on $N_{eval} = 1000$ points uniformly distributed within the whole time interval, and where $u_{eval}^i$ is the expected exact solution at $t = t_i$.
       }
\label{fig4}
\end{figure}

Nevertheless, for the following anharmonic equation,
\begin{equation}
         \frac {d^2u} {dt^2} + \omega_0^2  {u^3} = 0 ,
 \end{equation}
two training data were necessary (see Figure 13 in Paper 1 and left panel of Figure 3 in present paper). This is not a complete surprise as two conditions are necessary to integrate such second
order equation using analytic or classical numeric methods. Another option would be to solve an equivalent system of two first order differential equations, as done for the non linear
pendulum (Figure 12 in Paper 1). Indeed, in the latter case the NN is learning the solution and also the first order derivative on the whole time interval.

In this study, we propose another strategy. Indeed, we can choose the value of the first order time derivative evaluated at the first collocation
point, in order to constrain it to converge towards its true initial value $ \frac {du} {dt} (t = 0)$. Thus, the previous loss fonction $L_{data}$
that contains only the initial data value on $y (t = 0) = y_0$ can be modified to include a second term,
\begin{equation}
             L_{data}  = \left| {u_\theta} (t = 0 ) - u_0 \right|^2 + \omega_d \left| {u'_\theta} (t = 0 ) - u'_0 \right|^2 ,
 \end{equation}
where $u'_0 = \frac {du} {dt} (t = 0)$ and $ \omega_d$ is a new weight parameter not necessarily equal to one. This is an hybrid loss fonction as the
first term involves the training data and the second term the first collocation point (also used to evaluate the loss fonctions on equation residual
and total energy residual).
This is illustrated in Figure 3 for the anharmonic oscillator with initial conditions $u_0 = 1.5$ and $u'_0 = 0$, using also $\omega_0 = 15.5$ with
an integration over the time interval $t \in [0, 1]$. Indeed, we have obtained two PINN solutions choosing the following hyper parameters,
 $\eta = 1.1  \times 10^{-3}$, $\omega_{data} = 1$, $\omega_{\mathcal{F}} = 1  \times 10^{-5}$, and $\omega_E =  1  \times 10^{-6}$ for the
 two cases.  The NN architecture is made of 5 hidden layers with 32 neurons per layer. As explained above, in left panel of Figure 3 is plotted the converged
 PINN solution for the method without the improvement using two training data points (i.e. at $t = 0$, and $t = t_1 > 0$) and $N_c = 44$ collocation
 points. In right panel, one can see the second PINN solution for the method using the improvement (i.e. with the hybrid data loss function) with
 only one training data point and $N_c = 32$ collocation points. In the latter case, the chosen additional weight is $\omega_d = 1  \times 10^{-3}$.
The corresponding histories of the total loss and mean squared error ($MSE$) that are plotted in Figure 4, show
similar convergence during the training processes. Note that, this is important to also examine the $MSE$, as
it is a direct measure of the error contrary to the total loss that is a composite function in PINNs.

We can conclude that this first recipe using the hybrid data loss not only allows to reduce the training data set to the sole initial condition on $u$ (that is $u_0$,
as the initial derivative is imposed using the first collocation point), but it also allows to reduce the minimum number of collocation points.

\section{Applying a moving grid with a growing collocation interval}
\label{setup}

In Paper 1, it has been shown that when the differential equation is particularly stiff, the use of vanilla-PINNs requires a minimum number
of training data that is significantly higher than unity, and which is distributed over the time integration interval. This is indeed the case for Van Der Pol oscillator.

\subsection{Initial value problem - Van Der Pol oscillator }

 The Van Der Pol oscillator equation is
\begin{equation}
         \frac {d^2u} {dt^2} + \omega_0^2  {u} - \epsilon   \omega_0 (1 - u^2)  \frac {du} {dt} = 0,
 \end{equation}
where $ \omega_0$ is a normalized angular velocity, and $t \in [0, T]$.
Finally $\epsilon$ is a parameter having a value which determines the amplitude of a limit cycle in the phase space,
and consequently determines the stiffness of the equation.
PINN solutions obtained in Paper 1 for $\epsilon = 0.33, 1 $, and $5$, show that $3$, $7$, and $16$ training data points respectively were necessary for convergence
of the training process. One must note that, in this case the total energy is not conserved and therefore it is not possible to use the loss energy constraint.

\begin{figure}[!t]
\centering
 \includegraphics[scale=0.4]{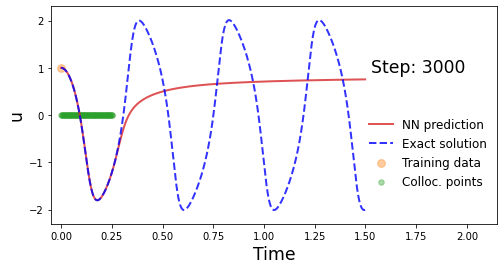}
 \includegraphics[scale=0.4]{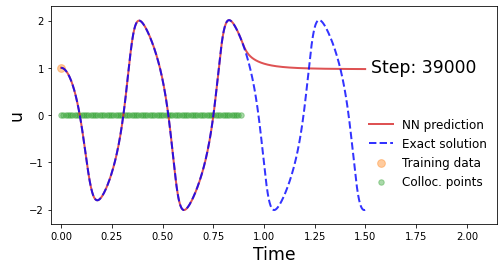}
 \includegraphics[scale=0.4]{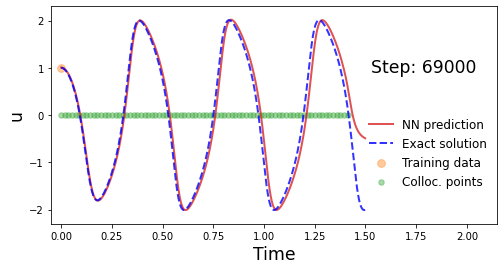}
 \includegraphics[scale=0.4]{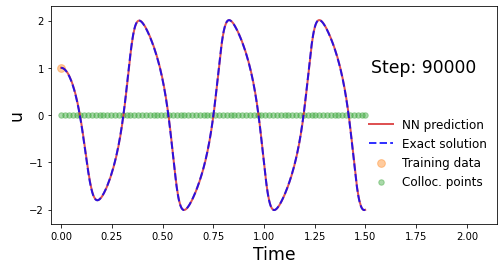}
  \caption{PINN solution for the Van Der Pol equation (see text), obtained using the moving grid procedure with a growing collocation interval. The different
  snapshots from top to bottom and left to right panels corresponds to the progression of the process as a function of the training step. The time
  at which the collocation points are defined are visible with the green circles.
         }
\label{fig5}
\end{figure}   

\begin{figure}[!t]
\centering
 \includegraphics[scale=0.6]{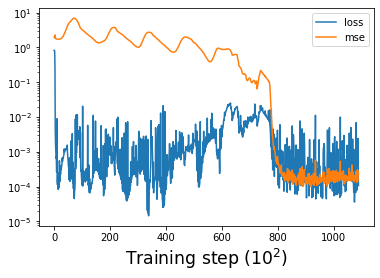}
   \caption{Histories of the total loss function $L$ and $MSE$ corresponding to the  case of the previous figure. }
\label{fig6}
\end{figure}   

We have found that vanilla-PINN, even when using the previous improvement strategy (with hybrid data loss), fails to train when the integration interval
covers a few periods of the oscillator. This is however not the case when the integration is done on a restricted time interval that is typically
smaller than one period. A strategy has been proposed to overcome PINNs failures in long time integration for PDE's, consisting in learning 
progressively sequence-to-sequence until the entire space-time solution is obtained (Krishnapriyan et al. 2021). In this strategy, the data set of collocation
points is progressively increased in order to finally invade the whole integration domain. New points are added along with the progression of the training process,
but they remain at fixed position in time once created. Hence, a considerably high number of points is necessary at the end of the training process.

We propose a variant of the latter strategy which requires a moderate and fixed number of collocation points. Indeed, we propose
to start the training process with $N_c$ collocation points uniformly distributed in a subdomain $[0, T_1]$ (with $T_1 < T$ and being typicaller
smaller than one oscillator period) of the whole time interval. In this way, our PINN algorithm can first learn the initial times. As the training progresses, we make the right
uppermost bound of the collocation data set interval moving towards the final time $T$. In other words, all the collocation points are
redefined (except the one imposed at $t = 0$) and moving while the collocation interval is expanding. The speed of progression must be
of course adapted to the rate of the training process. This is done manually in this work, but it can probably be adapted in a more
automatic way.

This other recipe is illustrated in Figures 5-6 for a moderately stiff case with $\epsilon =  1$. We have chosen $ \omega_0 = 15$ and integrated for $t \in \left[ 0,1.5 \right] $.
We also take the initial condition $u_0 = 1$ and zero initial derivative. The NN has $4$ hidden layers with $42$ neurons per layer. 
The number of training data and collocation points are $N_{data} = 1$ and $N_c = 80$ respectively.
The chosen weights are $\omega_{data} = 1$, $\omega_{d} = 0.1$, and $\omega_{\mathcal{F}} =  1  \times 10^{-4}$.
The learning rate is  $\eta = 7  \times 10^{-4}$. The initial upper bound $T_1$ is chosen as $T_1 = 0.2$. The different snapshots taken
at different training step clearly show how the solution is successfully learned along with the progression of the training process.
This is confirmed by the associated histories of the loss and $MSE$ plotted in Figure 6. The final convergence after $80 000$ epochs
is also clearly visible on the figure. Note that the exact solution is obtained by using a classical Runge-Kutta of order $4$ with $10000$
time steps uniformly distributed over the integration interval.

For completeness, we have also investigated a variant of the previous PINN modelling in which, instead of using the hybrid data loss
function to take into account the initial time derivative condition, we solve an equivalent system of two equations that is,
\begin{equation}
\centering
\left\{\begin{split}
    \frac {du_2} {dt} -   \frac {\omega_0} {\epsilon} u_1 = 0 , \\
   \frac {du_1} {dt}   -  \epsilon   \omega_0  (1 - \frac {u_1^2} {3}) u_1 +   \epsilon   \omega_0 u_2  = 0 
\end{split}\right.
 \end{equation}
The new expected variables are defined as $u_1 = u$, and $u_2 = u -  \frac {u^3} {3} - \frac {u'} {\epsilon   \omega_0}$ ($u'$ being the time derivative of $u$).
Contrary to another possible choice where $u_2 = u'$, this choice of variables called Lienard's transform is known to lead to values of $u_1$ and $u_2$ that have
similar magnitudes (see below). Note that the initial condition for $u' (t = 0) = 0$ is translated now into $u_2 (t = 0) = u_0 - u_0^3/3$,
that is $u_2 (t = 0) = 2/3$ for the case investigated.
The result of the corresponding PINN integration using our moving collocation grid procedure is illustrated in Figure 7.  
We have chosen $ \omega_0 = 15$ and integrated for $t \in \left[ 0,1.5 \right] $.
We also take the initial condition $u_0 = 1$ and zero initial derivative. The NN has $4$ hidden layers with $42$ neurons per layer. 
The number of training data and collocation points are $N_{data} = 2$ (one per solution variable) and $N_c = 90$ respectively.
The chosen weights are $\omega_{data} = 1$, and $\omega_{\mathcal{F}} =  9  \times 10^{-3}$.
The learning rate is  $\eta = 7  \times 10^{-4}$. Now, we do not need the hybrid data loss function, as the initial first order time derivative
is imposed via the training data set for $u_2$.

\begin{figure}[!t]
\centering
 \includegraphics[scale=0.4]{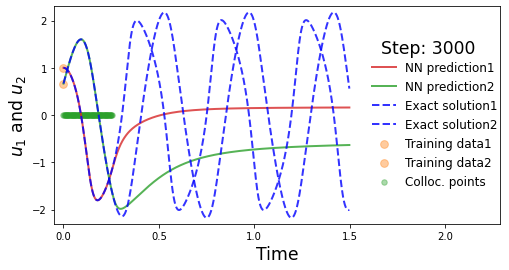}
 \includegraphics[scale=0.4]{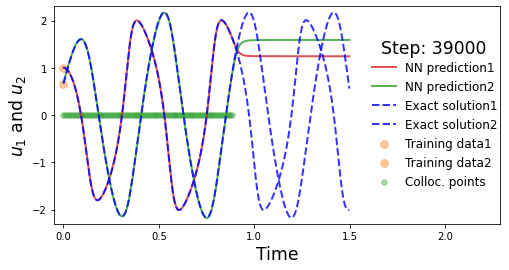}
 \includegraphics[scale=0.4]{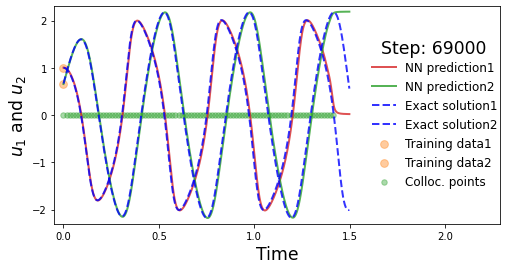}
 \includegraphics[scale=0.4]{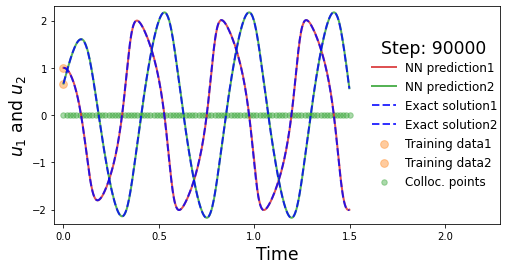}
  \caption{PINN solution of the Van der Pol equation solved as a two-dimensional system (see Equation 12) for the two variables $u_1$ and
  $u_2$. The moving grid with a growing collocation domain procedure is applied. The different
  snapshots from top to bottom and left to right panels corresponds to the progression of the process as a function of the training step. The time
  at which the collocation points are defined are visible with the green circles.
       }
\label{fig7}
\end{figure}   

However, one must note that our simple recipies have their own limitations. Indeed, when the stiffness of the system increases too much, one
is forced to find another strategy or an enlarged set of training data.

\subsection{Boundary value problem - steady-state solution of 1D convection diffusion equation}

\begin{figure}[!t]
\centering
 \includegraphics[scale=0.4]{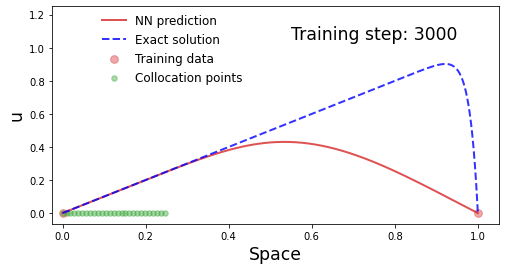}
 \includegraphics[scale=0.4]{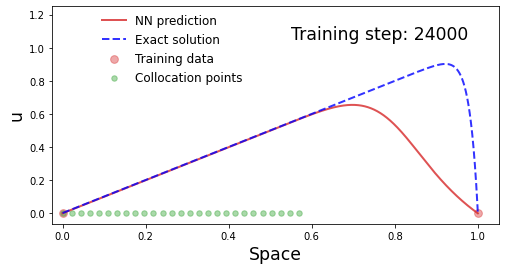}
 \includegraphics[scale=0.4]{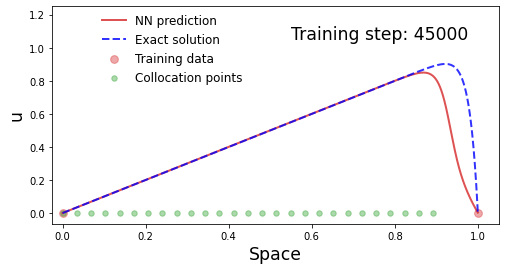}
 \includegraphics[scale=0.4]{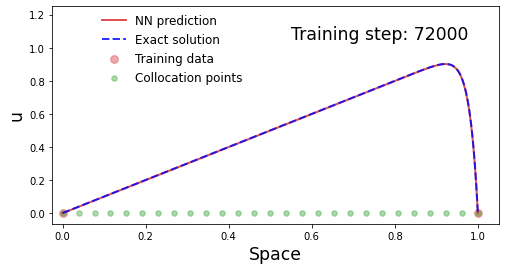}
  \caption{PINN solution of the steady-state 1D convection-diffusion equation (see Equation 13). The moving grid with a growing collocation interval procedure is applied. 
  The different snapshots from top to bottom and left to right panels corresponds to the progression of the process as a function of the training step. The space coordinate
  at which the collocation points are defined are visible with the green circles.
       }
\label{fig8}
\end{figure}   

\begin{figure}[!t]
\centering
 \includegraphics[scale=0.5]{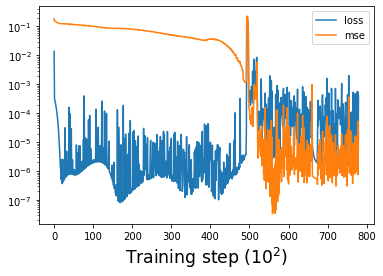}
   \caption{Histories of the total loss function $L$ and $MSE$ corresponding to the case of the previous figure. }
\label{fig9}
\end{figure}

In partial differential equations of high Reynolds convection-diffusion problems, imposed boundary conditions involve 
the formation of strongly localized boundary layers. This is a stiff system associated to BVP's which is considered to be numerically challenging.
In this section, we only focus
on the steady-state solution in one dimension. Hence, we consider the following equation,
\begin{equation}
  a  \frac {du} {dx} - \nu \frac {d^2u} {dx^2} - 1 = 0,
\end{equation}
where $a$ is a constant speed,  and $\nu$ is a viscosity parameter. The spatial integration domain
is for $x \in [0, 1]$, $x$ being a normalized spatial coordinate. 
The expected exact solution is,
\begin{equation}
  u (x) = x - \frac {e^{- R(1 -x)} - e^{- R}} {1 - e^{- R}} ,
\end{equation}
when the imposed boundary conditions are $u(0) = u (1) = 0$, with the definition of the Reynolds number $R = a/\nu$.

In order to illustrate our PINN solution using the previously described moving grid strategy with a growing collocation interval,
we have considered a high Reynolds case for $R = 50$ corresponding to $a = 1$ and $\nu = 0.02$. For this BVP, the two
boundary conditions are imposed via the training data set at $x = 0$ and $x = 1$. The results are plotted in Figures 8-9.
Indeed, one can clearly see that a modest number of only $N_c = 27$ collocation points is sufficient to obtain a good
training process. We have checked that, without such improvement recipe, more than $40$ collocation points are necessary.
Moreover, a classical numerical integration method also requires an even significantly larger number of points in the spatial domain for
uniform grid. For example, a finite-difference scheme requires a few points in order to resolve the quasi-singular layer which thickness is of order $1/R$.
For the PINN integration, we have chosen $3$ hidden layers with $42$ neurons per layer. 
The number of training data and collocation points are $N_{data} = 2$ and $N_c = 27$ respectively.
The chosen weights are $\omega_{data} = 1$, and $\omega_{\mathcal{F}} =  5  \times 10^{-1}$.
The learning rate is  $\eta = 1  \times 10^{-3}$

Finally, when the Reynolds number is higher, again our simple recipe shows its limitation and more sophisticated strategy
is required.

\section{Conclusion}

In this work, we have presented a few simple recipes with the aim to improve drawbacks of the vanilla-PINNs
for solving differential equations. This is indeed the case of stiff ODE's, requiring thus the use of a too large set
of data representing the prior knowledge of the solution. In other words, the sole knowledge of the initial/boundary
conditions is not sufficient. Modifying the total loss function in two ways, via an hybrid definition of the data loss, or
adding another partial loss associated to the conservation of the total energy (when it is possible) is a first
possible improvement. Another interesting idea consists in using a moving collocation grid with a growing interval
for the evaluation of the equation residual. In this way, the training process is made more progressive. More sophisticated
methods based on self-adaptive methods could be also developed (see McClenny \& Braga-Neto 2023,
Karniadakis et al. 2021, and Cuomo et al. 2022 for reviews).

Compared to classical numerical integration methods, PINNs still fail in terms of robustness because of
these failures inherent to the the training process. However, this technique using neural networks 
is relatively recent and many other improvements are probably expected in the future years. 
Nevertheless, the PINN formulation offers interesting advantages over classical methods. Indeed, it is a meshless method,
and once trained the solution can be quasi-instantaneously generated.

\begin{acknowledgments}{}
The author thanks Léo Baty (CERMICS, ENPC) for his help in Python programming.
Some of the Python codes used to make the figures are available from the Github repository at https://github.com/hubertbaty/PINNS-EDO2.
\end{acknowledgments}

\end{document}